\documentclass[12pt]{article}
\usepackage{amsmath}
\usepackage[english]{babel}
\usepackage{graphicx}
\usepackage{cite}
\title{\textbf{Analytical Approach for Calculating Chemotaxis Sensitivity Function}}
\author{\textit{Alex Vasilev}\\
\small Department of Theoretical Physics, Physics Faculty,\\
\small Taras Shevchenko National University of Kyiv,\\
\small 60 Volodymyrska Street, City of Kyiv, Ukraine, 01033\\
\small \textbf{e-mail:}\textit{vasilev@univ.kiev.ua}}
\date{}

\begin{document}
\maketitle

\begin{abstract}
We consider the chemotaxis problem for a one-dimensional system.
To analyze the interaction of bacteria and attractant we use a modified Keller-Segel model which accounts attractant absorption. To describe the system we use the chemotaxis sensitivity function, which characterizes nonuniformity of bacteria distribution. In particular, we investigate how the chemotaxis sensitivity function depends on the concentration of attractant at the boundary of the system. It is known that in the system without absorption the chemotaxis sensitivity function has a bell shape maximum. Here we show that attractant absorption and special boundary conditions for bacteria can cause the appearance of an additional maximum
in the chemotaxis sensitivity function. The value of this maximum  is determined
by the intensity of absorption.%
\end{abstract}

\section*{Introduction}
It is well known that when a bacterium like \textit{E. coli} is places in some substance (which is called \textit{attractant}) with the gradient of concentration then the bacterium moves toward the attractant gradient. This phenomenon is known as the \textit{chemotaxis} \cite{b01,b02,b03,b04,b05,b06}. Although there many interesting and significant results have been received in this area (for example, see \cite{b07,b08,b09,b10,b11,b12,b13,b14,b15,b16,b17,b18,b19,b20,b21,b22}), we are going to pay some attention to the process of the bacteria redistribution under the presence of attractant.

Frequently we don't need to know the exact spatial distribution of bacteria in the system. What we need is just some numerical characteristics that could be measured in an experiment. One of them is the \textit{chemotaxis sensitivity function}\cite{b06}. Namely, we will focus our attention on the one-dimensional system with attractant that is injected into the system at the left boundary. Technically it could be done by placing a capillary with attractant \cite{b06}. The system also contains bacteria which can interact with attractant. To investigate the system, we will use the methodological approach that was developed in \cite{b06}. In particular, our main goal will be the chemotaxis sensitivity function, which characterizes nonuniformity of the bacteria distribution. As it was shown in \cite{b06}, this function depends on the attractant concentration in a nontrivial way. Those results were received for the system with the linear distribution of attractant, that can be realized for the case when bacteria do not absorb attractant. Here we consider a more complex situation. But first of all we will make some comments about how we are to define the chemotaxis sensitivity function.

First of all, we consider a one-dimensional system whose spatial coordinate $x$ can change from $0$ to $L$ (i.e. $0\leq x\leq L$). Let it be that function $b(x)$ determines the spatial distribution of bacteria. We also assume that the system contains attractant and it is injected into the system at the left boundary with the help of some special capillary. The capillary, as it is supposed, is placed within the region $0\leq x\leq r_c$, where $r_c$ is the size of the capillary. At the right boundary of the system, the concentration of attractant is fixed at a lower level than it is at the left boundary. Then the concentration of bacteria should be the highest at the left boundary, within the region of the capillary.

For the described above one-dimensional system  the chemotaxis sensitivity function can be defined as follows \cite{b06}:
\begin{equation} \label{csf}
F=\frac{LP_{b}}{r_{c}}-1.
\end{equation}
Parameter $P_{b}$ in formula (\ref{csf}) is the probability to find a bacterium within the region $0\leq x\leq r_{c}$, and it is determined like this:
\begin{equation}
P_{b}=\frac{\int_{0}^{r_{c}}b(x)dx}{\int_{0}^{L}b(x)dx}.
\end{equation}
Actually, there in equation (\ref{csf}) we have the relation of the probability $P_{b}$ for a bacterium to be within the region of the capillary, to the probability $\frac{r_c}{L}$ for a bacterium to be within the region of the capillary if bacteria are distributed uniformly in the system. Thus, the chemotaxis sensitivity function is a numerical characteristics giving some notion of the bacteria distribution. If bacteria are distributed uniformly then $F=0$, and in the general case it can be of any sign. The greater the chemotaxis sensitivity function (by the modulus), the more nonuniform the bacteria distribution is.

It is also notable that in the limiting case when $r_c<<L$ we can rewrite the expression for the chemotaxis sensitivity function like this \cite{b06}:
\begin{equation}
F=\frac{Lb(0)}{\int_{0}^{L}b(x)dx}-1.
\end{equation}
Thus, to calculate the chemotaxis sensitivity function it is enough to know the total amount of bacteria in the system and the concentration of bacteria at the left boundary.

Next we consider the model which describes the one-dimensional system with bacteria and attractant. It is assumed that bacteria are redistributed according to the attractant gradient, and as well attractant is absorbed by bacteria. As it was mentioned above, to characterize such a system we will use the chemotaxis sensitivity function.

\section*{Basic Model}
To calculate the chemotaxis sensitivity function for the system with bacteria that absorb attractant, we use a model of the Keller-Segel kind \cite{k01,k02,k03}. As it is known, the classical Keller-Segel model is based on the following nonlinear partial differential equations \cite{k01}:
\begin{eqnarray}\label{eqA}
\partial_{t} a(t,\vec{r})=D_{a}\Delta a(t,\vec{r})+f_{1}(a,b),\\
\partial_{t} b(t,\vec{r})=D_{b}\Delta b(t,\vec{r})+f_{2}(a,b),\label{eqB}
\end{eqnarray}
where $\partial_{t}$ denotes the partial derivative on time $t$, $b(t,\vec{r})$ stands for the bacteria concentration, and $a(t,\vec{r})$ is the concentration of attractant. Parameters $D_{a}$ and $D_{b}$ are the diffusion coefficients. Function $f_{1}(a,b)$ accounts the absorption and the secretion of attractant, and function $f_{2}(a,b)$ defines the chemotactic flow of bacteria. If these functions are specified (as well as the boundary and initial conditions), then we can solve the system of equations (\ref{eqA})-(\ref{eqB}), at least in a numerical form \cite{m01,m02,m03,m04,m05,m06,m07,m08,m09}. As it was mentioned above, function $f_{1}(a,b)$ describes the attractant absorption (the attractant secretion will be accounted by the boundary conditions). Our assumptions concerning this function are as follows:
\begin{itemize}
\item the intensity of the attractant absorption is proportional to the bacteria density;
\item at low attractant concentration, the intensity of the attractant absorption is proportional to the concentration of attractant;
\item at high attractant concentration, the intensity of the attractant absorption does not depend on the attractant concentration.
\end{itemize}
All these allow us to consider function $f_{1}(a,b)$ to be like this:
\begin{equation}
f_{1}(a,b)=-k_{1}\frac{ab}{a_1+a},
\end{equation}
where $k_{1}$ and $a_{1}$ are phenomenological parameters of the model.
Our base assumption for function $f_{2}(a,b)$ is that the bacteria flux $j_{b}$ is determined by the bacteria concentration, its gradient, and the gradient of the attractant. In particular, we use the following formula for the bacteria flux:
\begin{equation}\label{flux}
j_{b}=-D_{b}\nabla b+b\varphi(a)\nabla a.
\end{equation}
The first term in equation (\ref{flux}) determines the flow of bacteria due to diffusion, and thus $D_{b}$ stands for the diffusion coefficient. The second term determines the bacteria flow caused by the inhomogeneity of the attractant distribution. It is supposed that this particular term is proportional to the bacteria concentration and also to the gradient of attractant. This term depends as well on the concentration of attractant in a nonlinear way. To account this dependence we use function $\varphi(a)$.

Thus, we can rewrite the equation that determines the temporal evolution of the bacteria distribution. In particular, we have the following:
\begin{equation}
\partial_{t}b=D_{b}\Delta b-\nabla\Big(b\varphi(a)\nabla a\Big).
\end{equation}
In the stationary case we get the equation, which ties the bacteria distribution and the attractant distribution:
\begin{equation}
D_{b}\Delta b-\nabla\Big(b\varphi(a)\nabla a\Big)=0.
\end{equation}
It can be reduced to the first order differential equation of the form
\begin{equation}
D_{b}\nabla b-b\varphi(a)\nabla a=0.
\end{equation}
For the one dimensional geometry (where $0\leq x\leq L$), this equation with the boundary condition
\begin{equation}
j_{b}\Big|_{z=0}=0
\end{equation}
(which means zero bacteria flux at the left boundary) produces the next formula determining the relation between the bacteria concentration $b(x)$ and the attractant concentration $a(x)$:
\begin{equation}\label{dens}
b(x)=A\exp\Big(\frac{1}{D_{b}}\int\varphi(a)da\Big).
\end{equation}
Integrand constant $A$ should be determined by another boundary condition for the bacteria distribution, which we will consider and discuss later.

To make some quantitative analysis we have to specify function $\varphi(a)$ (and by it, function $f_{2}(a,b)$). Here we take into account that the chemotaxis bacteria flow is proportional to the gradient of attractant at low attractant concentration, it is decreased (down to zero) with increasing the attractant concentration, and it is also proportional to the bacteria concentration. According to these we can present function $f_{2}(a,b)$ in the form
\begin{equation}
f_{2}(a,b)=k_{2}\nabla\Big(\frac{b\nabla a}{(a_2+a)^2}\Big)
\end{equation}
with phenomenological parameters $k_{2}$ and $a_{2}$. And thus function $\varphi(a)$ is as follows:
\begin{equation}
\varphi(a)=\frac{k_{2}}{(a_{2}+a)^2}.
\end{equation}
Then relation (\ref{dens}) between the bacteria and attractant concentrations can be rewritten like this:
\begin{equation}\label{bvsa}
b(x)=A\exp\Big(-\frac{k_{2}}{D_{b}}\frac{1}{a_{2}+a}\Big).
\end{equation}
Equation (\ref{bvsa}) gives the relation between the bacteria and attractant concentrations. We can compare it to the similar relation that was received in \cite{b06} for the particular case with the linear distribution of attractant in the system (the system without absorption). It is of the form \cite{b06}
\begin{equation}\label{bvsa2}
b(x)=A\Big(\frac{c_{1}+a(x)}{c_{2}+a(x)}\Big)^{N},
\end{equation}
where $A$ is a normalization constant (just the same as in equation (\ref{bvsa})), $c_{1,2}$ and $N$ are parameters of the model used in \cite{b06}.
Formally, these relations (\ref{bvsa}) and (\ref{bvsa2}) are different. Nevertheless, numerical estimations for the nondimensional parameter $N$ give that for the real systems $N>>1$. So if we perform the limiting transition $N\rightarrow \infty$, then we can get the following from formula (\ref{bvsa2}): 
\begin{equation}\label{bvsa3}
b(x)\approx A\exp\Big(-N\frac{c_2-c_1}{c_2+a(x)}\Big).
\end{equation}
We see that if $c_2=a_2$ and $k_2=D_{b}N(c_2-c_1)$, then equations (\ref{bvsa}) and (\ref{bvsa3}) determine the same dependencies. To estimate and compare dependencies that are given by equations (\ref{bvsa}) and (\ref{bvsa2}), we use the following values for the parameters (according to the data in \cite{b06}): $N\approx38.56$ and $c_2/c_1\approx 166.67$. Fig.~1 presents the bacteria concentration as a function of the attractant concentration, which is calculated according to formulae (\ref{bvsa}) and (\ref{bvsa2}). For the attractant concentration, it is taken that $a=a_2\cdot 10^p$. As we can easily see from Fig.~1, formulae (\ref{bvsa}) and (\ref{bvsa2}) give actually the same dependencies.
%%%%%%%%%%%%%%%%%%%%%%%%%%%%%%%%%%%%%%%%%%%%%%%%%%%%%%%%%%%%%%%%%%%%%%%%%%%%%%%%%%%%%%%%%%%%%%%%%%%%%%%%%%%%%%%%%%%%%%%%%
%%%%%                                                 Figure 1
%%%%%%%%%%%%%%%%%%%%%%%%%%%%%%%%%%%%%%%%%%%%%%%%%%%%%%%%%%%%%%%%%%%%%%%%%%%%%%%%%%%%%%%%%%%%%%%%%%%%%%%%%%%%%%%%%%%%%%%%%
\begin{figure}
\center{\resizebox{0.75\columnwidth}{!}{%
  \includegraphics[angle=0]{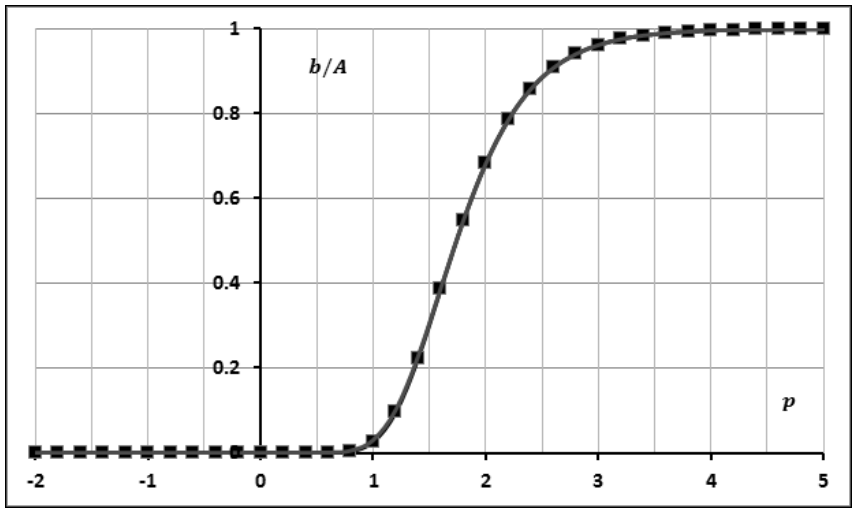}}
} \caption{The bacteria concentration as a function of the attractant concentration. It is taken that $a=a_2\cdot 10^p$. The solid line shows the dependence by formula (\ref{bvsa}). The points (squares) correspond to the dependence that is presented with formula (\ref{bvsa2}). It is also taken that $N\approx 38.56$ and $c_2/c_1\approx 166.67$}
\end{figure}
%%%%%%%%%%%%%%%%%%%%%%%%%%%%%%%%%%%%%%%%%%%%%%%%%%%%%%%%%%%%%%%%%%%%%%%%%%%%%%%%%%%%%%%%%%%%%%%%%%%%%%%%%%%%%%%%%%%%%%%%%
%%%%%                                                End of Figure 1
%%%%%%%%%%%%%%%%%%%%%%%%%%%%%%%%%%%%%%%%%%%%%%%%%%%%%%%%%%%%%%%%%%%%%%%%%%%%%%%%%%%%%%%%%%%%%%%%%%%%%%%%%%%%%%%%%%%%%%%%%
\section*{Attractant Distribution}
Taking into account previous results, we get the following differential equation for the stationary distribution of  attractant:
\begin{equation}
D_{a}\Delta a(x)-k_{1}A\exp\Big(-\frac{k_{2}}{D_{b}}\frac{1}{a_{2}+a}\Big)\frac{a(x)}{a_{1}+a(x)}=0.
\end{equation}
It should be supplemented with some boundary conditions for the attractant concentration, and with an additional condition for the bacteria distribution function. We consider the boundary conditions for the attractant distribution when the attractant concentration is fixed at the boundaries. This means that
\begin{eqnarray}
a(x=0)=C_{0},\\
a(x=L)=C_{1},
\end{eqnarray}
and parameters $C_{0}$ and $C_{1}$ (we assume that $C_{0}\geq C_{1}$) are given.

Our next step deals with redefining some parameters. In particular, for the sake of simplicity we use the substitutions $x=Lz$ and $a(x)=a_{2}s(z)$. Then we get the following equation:
\begin{equation}\label{main}
s''(z)-\alpha A\exp\Big(-\frac{\beta}{1+s(z)}\Big)\frac{s(z)}{\lambda+s(z)}=0,
\end{equation}
where we have used parameters $\alpha=\frac{k_{1}L^{2}}{D_{a}a_{2}}$, $\beta=\frac{k_{2}}{D_{b}a_{2}}$ and $\lambda=\frac{a_{1}}{a_{2}}$. The boundary conditions are transformed to these:
\begin{eqnarray}\label{bc1}
s(z=0)=\frac{C_{0}}{a_{2}}\equiv\gamma_{0},\\
\label{bc2}
s(z=1)=\frac{C_{1}}{a_{2}}\equiv\gamma_{1}.
\end{eqnarray}
The chemotaxis sensitivity function $F$ in this case is determined by the relation
\begin{equation}\label{SF}
F=\frac{b(0)}{\int_{0}^{1}b(z)dz}-1,
\end{equation}
where the bacteria distribution is given by the expression
\begin{equation}\label{dens2}
b(z)=A\exp\Big(-\frac{\beta}{1+s(z)}\Big).
\end{equation}
Thus, to solve the problem and find the value of the chemotaxis sensitivity function (basing on some additional restriction applied for the bacteria distribution function $b(z)$), we have to specify constant $A$ in equation (\ref{dens2}), solve
then equation (\ref{main}) with boundary conditions (\ref{bc1}) and (\ref{bc2}), and after that calculate the chemotaxis sensitivity function $F$ according to relation (\ref{SF}).
\section*{Chemotaxis Sensitivity Function}
Next we consider the chemotaxis sensitivity function and, in particular, let's clear out
how it depends on the attractant concentration at the left boundary of the system. It is understood that the chemotaxis sensitivity function
\begin{equation}
F=\frac{b(0)}{\int_{0}^{1}b(z)dz}-1=\frac{\exp\Big(-\frac{\beta}{1+\gamma_{0}}\Big)}{\int_{0}^{1}\exp\Big(-\frac{\beta}{1+s(z)}\Big)dz}-1,
\end{equation}
and it formally doesn't depend on $A$. Nevertheless, the solution for the bacteria distribution $s(z)$ is determined by equation (\ref{main}), which contains parameter $A$. So the chemotaxis sensitivity function depends implicitly on how we determine $A$. It in turns depends on the restriction we apply for the concentration $b(z)$ of bacteria. Here we will consider three regimes that specifies the distribution of bacteria:
\begin{itemize}
\item
the concentration of bacteria at the right boundary is fixed;
\item
the total amount (or mass) of the bacteria in the system is fixed;
\item
the concentration of bacteria at the right boundary is changed with changing the attractant concentration, to supply parameter $A$ to be fixed.
\end{itemize}
Fixing the bacteria concentration at the right boundary $b(1)=B_{1}$  gives the restriction
\begin{equation}
A\exp\Big(-\frac{\beta}{1+\gamma_{1}}\Big)=B_{1}.
\end{equation}
If the value of parameter $B_{1}$ is given, then to solve the problem we have to solve equation (\ref{main}), which in this case is transformed to the following:
\begin{equation}
s''(z)-\alpha B_{1}\exp\Big(\frac{\beta}{1+\gamma_{1}}\Big)\exp\Big(-\frac{\beta}{1+s(z)}\Big)\frac{s(z)}{\lambda+s(z)}=0.
\end{equation}
Then, knowing the distribution $s(z)$, we calculate the chemotaxis sensitivity function according to equation (\ref{SF}). Fig.~2 illustrates how the chemotaxis sensitivity function depends on the concentration of attractant at the left boundary of the system. In particular, we take $\gamma_0=10^p$ and $\gamma_1=\xi\cdot\gamma_0$, where $\xi=0.75$ is fixed, and parameter $p$ changes from $-3$ to $3$. It is also taken $\lambda=10$ and $\beta=38.56$.
%%%%%%%%%%%%%%%%%%%%%%%%%%%%%%%%%%%%%%%%%%%%%%%%%%%%%%%%%%%%%%%%%%%%%%%%%%%%%%%%%%%%%%%%%%%%%%%%%%%%%%%%%%%%%%%%%%%%%%%%%%%%%
%%%%%%%%%%%%%%%%                     Figure 2
%%%%%%%%%%%%%%%%%%%%%%%%%%%%%%%%%%%%%%%%%%%%%%%%%%%%%%%%%%%%%%%%%%%%%%%%%%%%%%%%%%%%%%%%%%%%%%%%%%%%%%%%%%%%%%%%%%%%%%%%%%%%%
\begin{figure}
\center{\resizebox{0.75\columnwidth}{!}{%
  \includegraphics[angle=0]{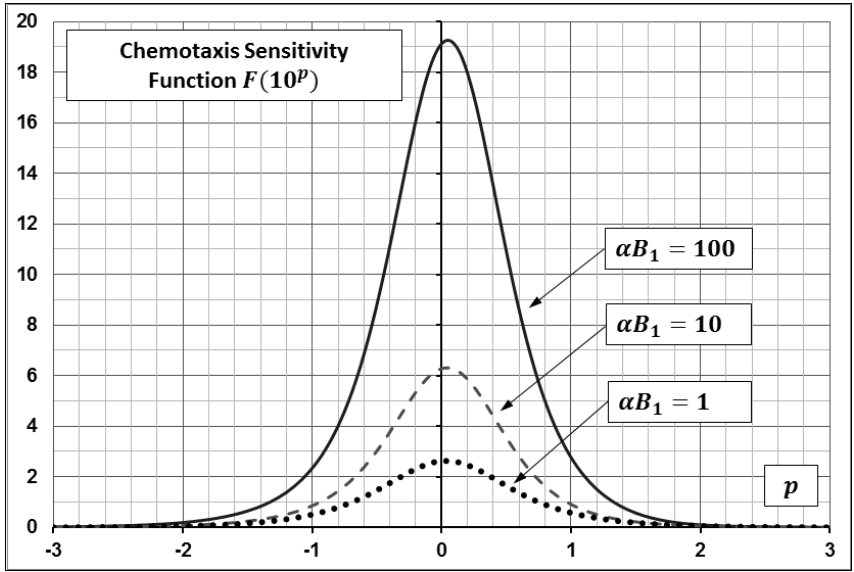}}
} \caption{The dependence of the chemotaxis sensitivity function on the concentration of attractant at the left boundary. It is taken that $\gamma_0=10^p$, $\gamma_1=\xi\cdot\gamma_0$, $\lambda=10$, $\beta=38.56$. The dotted line is for the value $\alpha B_1=1$, the dashed line is for the value $\alpha B_1=10$, and the solid line is for the value $\alpha B_1=100$}
\end{figure}
%%%%%%%%%%%%%%%%%%%%%%%%%%%%%%%%%%%%%%%%%%%%%%%%%%%%%%%%%%%%%%%%%%%%%%%%%%%%%%%%%%%%%%%%%%%%%%%%%%%%%%%%%%%%%%%%%%%%%%%%%%%%%
%%%%%%%%%%%%%%%%                     End of Figure 2
%%%%%%%%%%%%%%%%%%%%%%%%%%%%%%%%%%%%%%%%%%%%%%%%%%%%%%%%%%%%%%%%%%%%%%%%%%%%%%%%%%%%%%%%%%%%%%%%%%%%%%%%%%%%%%%%%%%%%%%%%%%%%
As we can see, dependence of the chemotaxis sensitivity function on the attractant concentration at the left boundary (more precisely, on parameter $p$) has a bell shape maximum. The value of the maximum depends on the parameters of the model. But the matter of fact is that the maximum exists, and that it is the only maximum.

Fixing the total amount of bacteria in the system $\int_{0}^{1}b(z)=B_{2}$ gives the following restriction:
\begin{equation}\label{R3}
A\int_{0}^{1}\exp\Big(-\frac{\beta}{1+s(z)}\Big)dz=B_{2}.
\end{equation}
In this case, to find the chemotaxis sensitivity function we actually have to solve a system of equations. The first one is equation (\ref{main}). It contains parameter $A$. On the other hand, this parameter is to satisfy relation (\ref{R3}), which in turns contains the solution $s(z)$ of equation (\ref{main}). Numerical calculations for this problem show that dependence of the chemotaxis sensitivity function on the concentration of attractant (at the left boundary) is the same qualitatively as in the previous case (when we fix the bacteria concentration at the right boundary). Fig.~3 compares these two cases. It contains plots for the chemotaxis sensitivity functions that were calculated a) with the fixed bacteria concentration at the right boundary, and b) with the fixed total amount of bacteria in the system.
%%%%%%%%%%%%%%%%%%%%%%%%%%%%%%%%%%%%%%%%%%%%%%%%%%%%%%%%%%%%%%%%%%%%%%%%%%%%%%%%%%%%%%%%%%%%%%%%%%%%%%%%%%%%%%%%%%%%%%%%%%%%%
%%%%%%%%%%%%%%%%                     Figure 3
%%%%%%%%%%%%%%%%%%%%%%%%%%%%%%%%%%%%%%%%%%%%%%%%%%%%%%%%%%%%%%%%%%%%%%%%%%%%%%%%%%%%%%%%%%%%%%%%%%%%%%%%%%%%%%%%%%%%%%%%%%%%%
\begin{figure}
\center{\resizebox{0.75\columnwidth}{!}{%
  \includegraphics[angle=0]{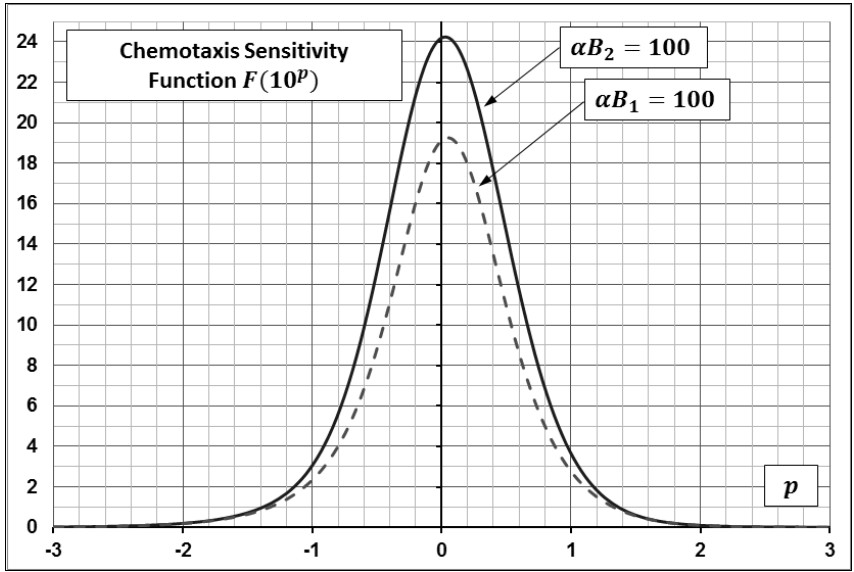}}
} \caption{The dependence of the chemotaxis sensitivity function on the concentration of attractant at the left boundary. It is taken that $\gamma_0=10^p$, $\gamma_1=\xi\cdot\gamma_0$, $\lambda=10$, $\beta=38.56$. The dashed line is for the value $\alpha B_1=10$ (the bacteria concentration at the right boundary is fixed), and the solid line is for the value $\alpha B_2=100$ (the total amount of bacteria is fixed)}
\end{figure}
%%%%%%%%%%%%%%%%%%%%%%%%%%%%%%%%%%%%%%%%%%%%%%%%%%%%%%%%%%%%%%%%%%%%%%%%%%%%%%%%%%%%%%%%%%%%%%%%%%%%%%%%%%%%%%%%%%%%%%%%%%%%%
%%%%%%%%%%%%%%%%                     End of Figure 3
%%%%%%%%%%%%%%%%%%%%%%%%%%%%%%%%%%%%%%%%%%%%%%%%%%%%%%%%%%%%%%%%%%%%%%%%%%%%%%%%%%%%%%%%%%%%%%%%%%%%%%%%%%%%%%%%%%%%%%%%%%%%%

The third scenario is when we change the concentration of bacteria at the right boundary synchronously with changing the attractant concentration at the left boundary. In particular, we take the following boundary condition for the bacteria concentration:
\begin{equation}
b(1)=B_{3}\exp(-\frac{\beta}{1+\gamma_{1}}).
\end{equation}
This gives the condition $A=B_{3}$ for solving equation (\ref{main}). Fig.~4 shows how the chemotaxis sensitivity function looks like in this case. The most important thing is that it may have two maximums. In particular, increasing the value of the product $\alpha B_{3}$ leads to appearing of an additional maximum at high concentration of attractant. So it is clear that this effect is caused by the attractant absorption. It is also notable that the way we take the boundary condition for bacteria is important. Fig.~5 illustrates how the total amount of bacteria and the bacteria concentration at the right boundary change with changing the attractant concentration at the left boundary of the system. All these characteristics are normalized for the $B_{3}$ constant. For the sake of simplicity, Fig.~5 also contains the plot for the chemotaxis sensitivity function. And what we can see is that the region of the second additional maximum coincides with the region where the bacteria concentration is increased.
%%%%%%%%%%%%%%%%%%%%%%%%%%%%%%%%%%%%%%%%%%%%%%%%%%%%%%%%%%%%%%%%%%%%%%%%%%%%%%%%%%%%%%%%%%%%%%%%%%%%%%%%%%%%%%%%%%%%%%%%%%%%%
%%%%%%%%%%%%%%%%                     Figure 4
%%%%%%%%%%%%%%%%%%%%%%%%%%%%%%%%%%%%%%%%%%%%%%%%%%%%%%%%%%%%%%%%%%%%%%%%%%%%%%%%%%%%%%%%%%%%%%%%%%%%%%%%%%%%%%%%%%%%%%%%%%%%%
\begin{figure}
\center{\resizebox{0.75\columnwidth}{!}{%
  \includegraphics[angle=0]{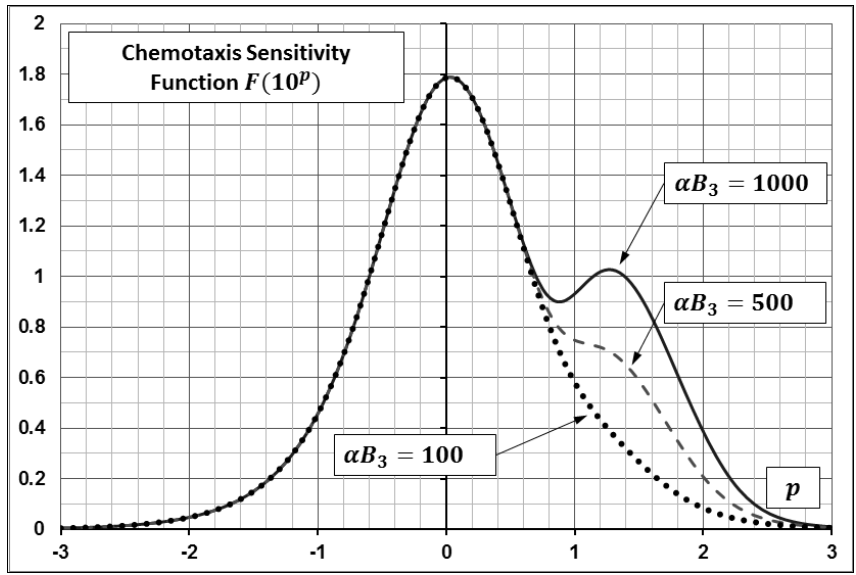}}
} \caption{The dependence of the chemotaxis sensitivity function on the concentration of attractant at the left boundary. It is taken that $\gamma_0=10^p$, $\gamma_1=\xi\cdot\gamma_0$, $\lambda=10$, $\beta=38.56$. The dotted line is for the value $\alpha B_{3}=100$, the dashed line is for the value $\alpha B_3=500$, and the solid line is for the value $\alpha B_{3}=1000$}
\end{figure}
%%%%%%%%%%%%%%%%%%%%%%%%%%%%%%%%%%%%%%%%%%%%%%%%%%%%%%%%%%%%%%%%%%%%%%%%%%%%%%%%%%%%%%%%%%%%%%%%%%%%%%%%%%%%%%%%%%%%%%%%%%%%%
%%%%%%%%%%%%%%%%                     End of Figure 4
%%%%%%%%%%%%%%%%%%%%%%%%%%%%%%%%%%%%%%%%%%%%%%%%%%%%%%%%%%%%%%%%%%%%%%%%%%%%%%%%%%%%%%%%%%%%%%%%%%%%%%%%%%%%%%%%%%%%%%%%%%%%%

%%%%%%%%%%%%%%%%%%%%%%%%%%%%%%%%%%%%%%%%%%%%%%%%%%%%%%%%%%%%%%%%%%%%%%%%%%%%%%%%%%%%%%%%%%%%%%%%%%%%%%%%%%%%%%%%%%%%%%%%%%%%%
%%%%%%%%%%%%%%%%                     Figure 5
%%%%%%%%%%%%%%%%%%%%%%%%%%%%%%%%%%%%%%%%%%%%%%%%%%%%%%%%%%%%%%%%%%%%%%%%%%%%%%%%%%%%%%%%%%%%%%%%%%%%%%%%%%%%%%%%%%%%%%%%%%%%%
\begin{figure}
\center{\resizebox{0.75\columnwidth}{!}{%
  \includegraphics[angle=0]{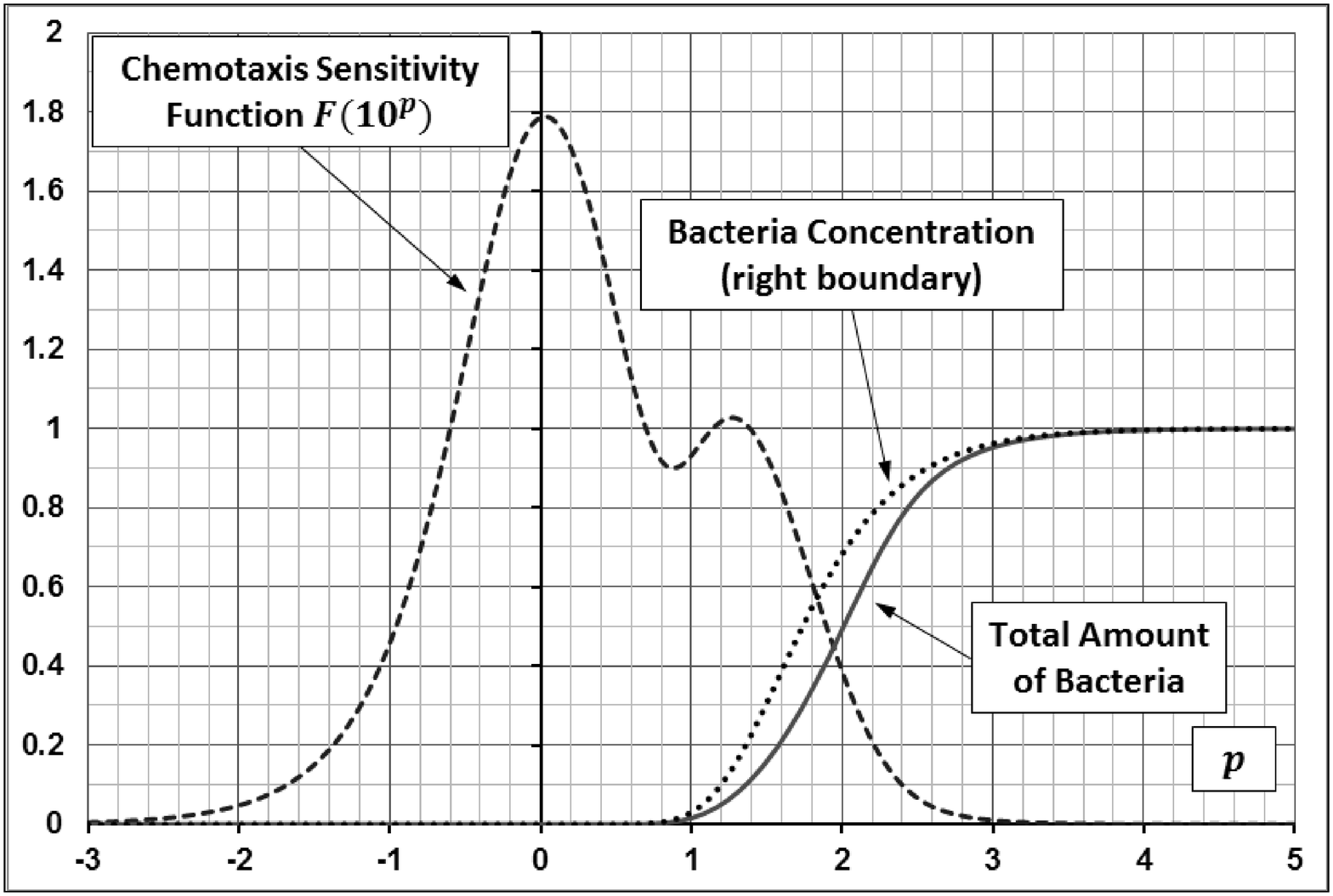}}
} \caption{The chemotaxis sensitivity function and the bacteria total amount. It is taken that $\gamma_0=10^p$, $\gamma_1=\xi\cdot\gamma_0$, $\lambda=10$, $\beta=38.56$, the value $\alpha B_{3}=1000$. The dotted line shows the concentration of bacteria at the right boundary, the solid line shows the total amount of bacteria in the system, and the dashed line demonstrates how the chemotaxis sensitivity function depends on the attractant concentration at the left boundary of the system. The bacteria concentration is taken without multiplier $B_{3}$ (which is a constant)}
\end{figure}
%%%%%%%%%%%%%%%%%%%%%%%%%%%%%%%%%%%%%%%%%%%%%%%%%%%%%%%%%%%%%%%%%%%%%%%%%%%%%%%%%%%%%%%%%%%%%%%%%%%%%%%%%%%%%%%%%%%%%%%%%%%%%
%%%%%%%%%%%%%%%%                     End of Figure 5
%%%%%%%%%%%%%%%%%%%%%%%%%%%%%%%%%%%%%%%%%%%%%%%%%%%%%%%%%%%%%%%%%%%%%%%%%%%%%%%%%%%%%%%%%%%%%%%%%%%%%%%%%%%%%%%%%%%%%%%%%%%%%
\section*{Discussion}
So, when we fix the bacteria concentration at the right boundary or the total amount of bacteria in the system, then the chemotaxis sensitivity function has a bell shape maximum. It has a quite obvious explanation \cite{b06}. Matter of fact is that when we increase the attractant concentration then the gradient of the attractant concentration is increased as well. Due to the increasing of the gradient, the bacteria distribution becomes more nonuniform, and thus the chemotaxis sensitivity function is increased. But at high levels of the attractant concentration the bacteria reaction on the attractant gradient is decreased. In other words, bacteria "don't feel" the gradient when the attractant concentration is significant. Thus, the bacteria distribution becomes more uniform and the chemotaxis sensitivity function is decreased. From physiological point of view it can be explained in the way a bacterium behaves in the system with attractant. What we know is that every bacterium has receptors which can interact with attractant (for example, see \cite{b06} and the references in it). The amount of the receptors that are in interaction with attractant determines the methylation level of the bacterium \cite{b06}. Any bacterium moves straight with a constant velocity. But from time to time it changes the direction of its motion. These acts are called tumbles. It is generally accepted that the new direction of motion is selected randomly. And the frequency of tumbles depends on the methylation level of the bacterium. The greater the methylation level, the smaller the tumble frequency is. Actually, this is the simplified mechanism of how bacteria behave within the system with attractant. And it is clear that if the attractant concentration is high enough, then the methylation level can be at the highest possible level. Thus, bacteria can't react on the changes of the attractant concentration \cite{b06}.
%%%%%%%%%%%%%%%%%%%%%%%%%%%%%%%%%%%%%%%%%%%%%%%%%%%%%%%%%%%%%%%%%%%%%%%%%%%%%%%%%%%%%%%%%%%%%%%%%%%%%%%%%%%%%%%%%%%%%%%%%%%%%
%%%%%%%%%%%%%%%%                     Figure 6
%%%%%%%%%%%%%%%%%%%%%%%%%%%%%%%%%%%%%%%%%%%%%%%%%%%%%%%%%%%%%%%%%%%%%%%%%%%%%%%%%%%%%%%%%%%%%%%%%%%%%%%%%%%%%%%%%%%%%%%%%%%%%
\begin{figure}
\center{\resizebox{0.75\columnwidth}{!}{%
  \includegraphics[angle=0]{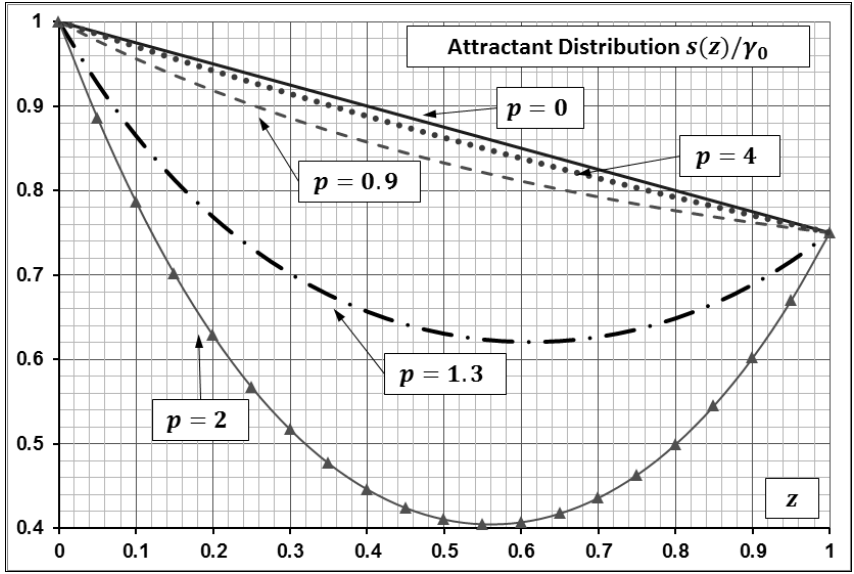}}
} \caption{The attractant distribution $s(z)/\gamma_{0}$ for the different values of parameter $p$ (it is taken that $\gamma_0=10^p$, $\gamma_1=\xi\cdot\gamma_0$, $\lambda=10$, $\beta=38.56$, the value $\alpha B_{3}=1000$): the solid line is for the value $p=0$, the dashed line is for the value $p=0.9$, the dash-dotted line is for the value $p=1.3$, the solid line with triangular markers is for the value $p=2$, and the dotted line is for the value $p=4$}
\end{figure}
%%%%%%%%%%%%%%%%%%%%%%%%%%%%%%%%%%%%%%%%%%%%%%%%%%%%%%%%%%%%%%%%%%%%%%%%%%%%%%%%%%%%%%%%%%%%%%%%%%%%%%%%%%%%%%%%%%%%%%%%%%%%%
%%%%%%%%%%%%%%%%                     End of Figure 6
%%%%%%%%%%%%%%%%%%%%%%%%%%%%%%%%%%%%%%%%%%%%%%%%%%%%%%%%%%%%%%%%%%%%%%%%%%%%%%%%%%%%%%%%%%%%%%%%%%%%%%%%%%%%%%%%%%%%%%%%%%%%%
%%%%%%%%%%%%%%%%%%%%%%%%%%%%%%%%%%%%%%%%%%%%%%%%%%%%%%%%%%%%%%%%%%%%%%%%%%%%%%%%%%%%%%%%%%%%%%%%%%%%%%%%%%%%%%%%%%%%%%%%%%%%%
%%%%%%%%%%%%%%%%                     Figure 7
%%%%%%%%%%%%%%%%%%%%%%%%%%%%%%%%%%%%%%%%%%%%%%%%%%%%%%%%%%%%%%%%%%%%%%%%%%%%%%%%%%%%%%%%%%%%%%%%%%%%%%%%%%%%%%%%%%%%%%%%%%%%%
\begin{figure}
\center{\resizebox{0.75\columnwidth}{!}{%
  \includegraphics[angle=0]{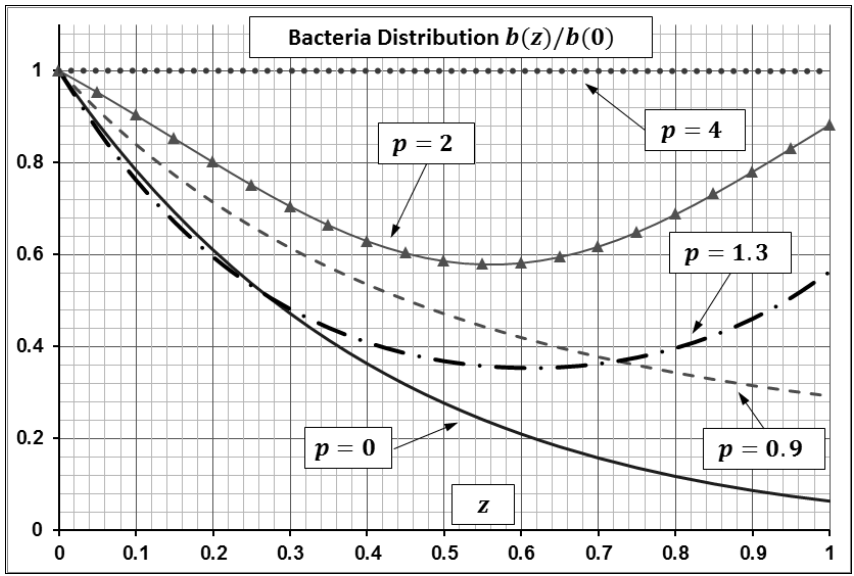}}
} \caption{The bacteria distribution $b(z)/b(0)$ for the different values of parameter $p$ (it is taken that $\gamma_0=10^p$, $\gamma_1=\xi\cdot\gamma_0$, $\lambda=10$, $\beta=38.56$, the value $\alpha B_{3}=1000$): the solid line is for the value $p=0$, the dashed line is for the value $p=0.9$, the dash-dotted line is for the value $p=1.3$, the solid line with triangular markers is for the value $p=2$, and the dotted line is for the value $p=4$}
\end{figure}
%%%%%%%%%%%%%%%%%%%%%%%%%%%%%%%%%%%%%%%%%%%%%%%%%%%%%%%%%%%%%%%%%%%%%%%%%%%%%%%%%%%%%%%%%%%%%%%%%%%%%%%%%%%%%%%%%%%%%%%%%%%%%
%%%%%%%%%%%%%%%%                     End of Figure 7
%%%%%%%%%%%%%%%%%%%%%%%%%%%%%%%%%%%%%%%%%%%%%%%%%%%%%%%%%%%%%%%%%%%%%%%%%%%%%%%%%%%%%%%%%%%%%%%%%%%%%%%%%%%%%%%%%%%%%%%%%%%%%

In our model the effect of the bell shape chemotaxis sensitivity function can be explained if we take into account the relation between the bacteria and attractant concentrations (see equation (\ref{bvsa}) and Fig.~1). It gives that when the attractant concentration is high then the bacteria concentration is at the saturation level, and further increasing the attractant concentration doesn't change the bacteria concentration. And thus, presence of the gradient of the attractant concentration is not tested by bacteria.

The situation with two maximums of the chemotaxis sensitivity function is explained in the way that when we change the bacteria concentration at the right boundary, then we actually change the total amount of bacteria in the system. If bacteria didn't absorb attractant then the change of their total amount wouldn't affect the attractant distribution, and it in turns wouldn't change the bacteria distribution. Due to the attractant absorption, increasing the total amount of bacteria in the system changes the attractant distribution. The situation is illustrated in Fig.~6, where the plots are presented for the attractant distribution $s(z)/\gamma_{0}$ for different values of parameter $p$. In particular, there we can see that at the value $p=0$ the distribution is almost linear. With further increasing the value of parameter $p$, the distribution becomes more nonlinear, but then it comes back to the almost linear trend. Say, for the value $p=4$ (the solid line in Fig.~6) the attractant distribution is very close to the distribution under the value $p=0$ (the dotted line in Fig.~6).

The bacteria distribution is changed in a slightly different way. Fig.~7 contains plots for the bacteria distribution $b(z)/b(0)$ in the system for some values of parameter $p$. For example, at the value $p=0$ (the solid line in Fig.~7) it is decreased monotonously from the left boundary to the right boundary. With increasing the value of parameter $p$ the slope of the curve is decreased (in Fig.~7, see the dashed line for $p=0.9$) simultaneously with appearing of the minimum in the distribution (in Fig.~7, see the dash-dotted line for $p=1.3$ and the solid line with triangular markers for $p=2$). Then decreasing the value of the minimum gives the almost homogeneous distribution of bacteria in the system (in Fig.~7, see the dotted line for $p=4$).

\end{document}